\documentclass[universe,article,accept,pdftex,moreauthors]{Definitions/mdpi} 
\usepackage{amsmath}
\usepackage{graphicx}
\usepackage{inputenc}
\usepackage{bm}
\usepackage{multirow}
\usepackage{float}
\usepackage{url}
\usepackage{mathrsfs}
\usepackage{physics}
\usepackage{comment}
\usepackage{booktabs}
\usepackage{comment}
\usepackage{wasysym}
\long\def\OFF#1{}

\firstpage{1} 
\pubvolume{1}
\issuenum{1}
\articlenumber{0}
\pubyear{2025}
\copyrightyear{2025}
\datereceived{ } 
\daterevised{ } 
\dateaccepted{ } 
\datepublished{ } 
\hreflink{https://doi.org/} 

\Title{Neutron Decay Anomaly and Its Effects on Neutron Star Properties}

\TitleCitation{Neutron Decay Anomaly and Its Effects on Neutron Star Properties}

\Author{H. C. Das *\orcidA{}, G. F. Burgio\orcidB{}}

\AuthorNames{Firstname Lastname, Firstname Lastname and Firstname Lastname}

\isAPAStyle{%
       \AuthorCitation{Lastname, F., Lastname, F., \& Lastname, F.}
         }{%
        \isChicagoStyle{%
        \AuthorCitation{Lastname, Firstname, Firstname Lastname, and Firstname Lastname.}
        }{
        \AuthorCitation{Das, H.C.; Burgio, G.F.}
        }
}

\address[1]{%
INFN Sezione di Catania, Via Santa Sofia 64, 95123 Catania, Italy}

\corres{\hangafter=1 \hangindent=1.05em \hspace{-0.82em} Correspondence: harishchandra.das@ct.infn.it}

\abstract{We investigate the effects of dark matter (DM) on neutron star (NS) properties using the neutron decay anomaly model within the relativistic mean-field (RMF) framework. Three nucleonic models (HCD0–HCD2) are developed, satisfying astrophysical constraints such as the maximum NS mass ($\geq 2M_\odot$), the NICER mass-radius limits, and the tidal deformability constraint from the GW170817 event. The equation of states of the NS admixed with DM (DMANS) are calculated by incorporating the self-interactions between them. The macroscopic properties, such as mass, radius, and tidal deformability of the NSs, are obtained for HCD models along with five others by varying self-interaction strength. By combining NS observations with scattering cross-section constraints from galaxy clusters, we explore model-dependent trends in the DM self-interaction parameter space. While the quantitative bounds may vary with hadronic model choice, our analysis offers insights into the interplay between DM interactions and NS observables within the RMF framework.
}

\keyword{Neutron Star; Dark Matter; Equation of State} 

\begin{document}

\section{Introduction}
\label{sec:intro}
A free neutron breaks into a proton, an electron, and an antineutrino via $\beta-$decay outside the nucleus. However, a still challenging and longstanding issue is the precise measurement of the neutron lifetime. Two well-known experiments, performed by the bottle method and the beam experiment, measured the lifetime of neutrons as $\tau_n = 878.4\pm 0.5$ s \cite{ParticleDataGroup_2024}, and $\tau_n = 888.0\pm 2.0$ s \cite{Czarnecki_2018} respectively. This discrepancy in measuring the lifetime with two experiments raises the question ``whether the measurement methods suffer from a systematic error or the theoretical description of the neutron decay is incomplete". Therefore, \citet{Fornal_2018} focused on resolving this discrepancy by assuming that this anomaly might be due to an incomplete theoretical description of neutron decay, which is referred to as the ``neutron decay anomaly" (NDA).

According to their proposed hypothesis, this discrepancy could be resolved if 1\% neutrons decay into dark matter (DM). Based on this, the available decay channels are $n\rightarrow \chi + \gamma$, $n \rightarrow \chi + \phi$, and $n \rightarrow \chi + e^- + e^+$ \cite{Fornal_2018, Baym_2018, Mckeen_2018}. However, the first and last decay channels are ruled out by \citet{Tang_2018, Sun_2018} from their experimental outcomes based on the ultracold neutron asymmetry experiment at Los Alamos National Laboratory. Therefore, in this study, we use the $n\rightarrow \chi + \phi$ channel, where the neutron decays to a spin-half DM with a unit baryon number $\chi$ and a light-dark boson $\phi$. Recently, this channel got considerable attention in the study of neutron stars (NS), as discussed in Refs. \cite{Motta_2018, Grienstein_2019, Husain_2022a, Strumia_2022, Shirke_2023, Routaray_2024}. However, the primary objective of this study is to place joint constraints on the free parameters of the DM using observational data from NSs, including mass, radius, and tidal deformability, alongside scattering cross-section constraints from different galaxy clusters. Unlike previous studies, the main motivation of our work is to provide a bound for the self-interaction strength parameter by incorporating scattering cross-section data from multiple galaxies.

The application of this model to a NS provides us with a unique scenario because of its huge neutron-rich environment. The in-situ formation of DM inside the NS might affect its micro- and macroscopic properties, which directly affect the equation of state (EOS). This is because decays of neutrons to DM significantly affect the chemical and $\beta-$equilibrium, which governs the stability of the NS. The signature of the creation of the DM inside the NS could be an indirect detection probe for DM by observing different NS properties derived from pulsar timing measurements, X-ray observations, and gravitational wave detections, which will provide robust constraints on the properties of DM.

Several mechanisms also contribute to the presence of DM, including the accretion of DM into NSs \cite{Bell_2021} and its inheritance during supernova explosions \cite{Ciarcelluti_2011}. Moreover, the possible existence of DM admixed NS can also be described either by (i) single-fluid model (interaction between DM and hadronic matter by boson exchange) \cite{Das_2020, Das_2020_curv, Das_2021_inspiral, Das_2021_fmode, Das_2021_GW19, Das_2021_Galaxy, Kumar_2022, Routaray_2023, Kumar_2025, Mu_2023, Sen_2021, Dutra_2022} or (ii) two-fluid model (interaction is mainly due to gravitation) \cite{ Gresham_2019, RafieiKarkevandi_2022, Routaray_2023, Rutherford_2023, Caballero_2024, Liu_2024}. Each model has its own significance, depending on the approach used to study the properties of NSs. 

Therefore, prior studies are necessary to further investigate the interaction between DM and nucleons inside NSs, particularly when considering a single-fluid system. Hence, in this study, we use this NDA model to explore several macroscopic properties such as mass, radius, and tidal deformability. Although several studies have used this approach \cite{Baym_2018, Mckeen_2018, Motta_2018, Grienstein_2019, Husain_2022a, Husain_2022b, Shirke_2023, Routaray_2024}, in this paper we will discuss the crucial effects of DM interaction, mainly its self-interacting strength, on the macroscopic properties of the NS, and try to put constraints on unknown parameters of DM with the help of NS observational and galaxy clusters data. The detailed discussions are given in the following sections. 
\section{Methodology}
\label{sec:method}
\subsection{RMF Model Lagrangian}
\label{subsec:RMFML}
In RMF model, the nuclear interactions are carried out by the exchange of different mesons such as $\sigma$, $\omega$, and $\rho$, including the self and cross-couplings between them. In this case, we take up to 4th-order and more significant terms crucial for nuclear matter and NSs. The Lagrangian density is given as \cite{Kumar_2020}
\begin{align}
{\cal L}_N & = \sum_{N=p,n} \bar\psi_N
\Bigg\{\gamma_{\mu}(i\partial^{\mu} - g_{\omega}\omega^{\mu}
-\frac{1}{2} g_{\rho}\vec{\tau}_N\!\cdot\!\vec{\rho}^{\,\mu})
-(m_N - g_{\sigma}\sigma)\Bigg\} \psi_N
\nonumber\\
&
+\frac{1}{2}\partial^{\mu}\sigma\,\partial_{\mu}\sigma
-\frac{1}{2}m_{\sigma}^{2}\sigma^2
-\frac{\kappa_3 g_\sigma m_\sigma^2 \sigma^3}{3! \ m_N}
-\frac{\kappa_4 g_\sigma^2 m_\sigma^2 \sigma^4}{4! \ m_N^2}
\nonumber \\
& 
-\frac{1}{4}\omega^{\mu\nu}\omega_{\mu\nu}
+\frac{1}{2}m_{\omega}^{2}\omega^{\mu}\omega_{\mu}
+\frac{\zeta_0}{4!}g_\omega^2
(\omega^{\mu}\omega_{\mu})^2
\nonumber\\
& 
-\frac{1}{4}\vec \rho^{\mu\nu}\!\cdot\!\vec \rho_{\mu\nu}
+\frac{1}{2}m_{\rho}^{2}\vec\rho^{\mu}\!\cdot\!\vec\rho_{\mu} 
+\Lambda_{\omega}g_{\omega}^2g_{\rho}^2(\omega^{\mu}\omega_{\mu})(\vec\rho^{\,\nu}\!\cdot\!\vec\rho_{\nu}),
\label{eq:rmf_lagrangian}
\end{align}
where $g_\sigma$ ($m_\sigma$), $g_\omega$ ($m_\omega$), and $g_\rho$ ($m_\rho$) are the coupling constants (masses) for the $\sigma$, $\omega$, and $\rho$ mesons respectively. $m_N$ is the nucleon mass taken as 939 MeV; $\kappa_3$, $\kappa_4$, and $\zeta_0$ are the self-interacting coupling constants of the $\sigma$ and $\omega$ mesons respectively; $\Lambda_\omega$ is the cross-coupling between $\omega-\rho$ meson. $\omega^{\mu\nu} (= \partial^\mu\omega^\nu-\partial^\nu\omega^\mu$) and $\vec \rho^{\mu\nu} (=\partial^\mu\vec\rho^{\,\nu}-\partial^\nu\vec\rho^{\,\mu}$) are the field strength tensors for the $\omega$ and $\rho$ mesons respectively. $\vec\tau_N$ is the Pauli matrix having value $+1$, and $-1$ for proton and neutron respectively. The masses of $\sigma$, $\omega$, and $\rho$ mesons are considered to be 500, 783, and 763 MeV, respectively.

The field equations for all mesons can be derived within the mean-field approximation from the above Lagrangian. The energy density can be derived from the energy-momentum tensor as follows \cite{Kumar_2020, Pradhan_2023}:
\begin{align}
\epsilon_N & = \sum_{N=p,n} \frac{1}{\pi^2}\int_{0}^{k_N} k^2 \sqrt{k^2+m^{*2}} \, dk +m_{\sigma}^2{\sigma}^2\Bigg(\frac{1}{2}+\frac{\kappa_{3}}{3!}\frac{g_\sigma\sigma}{m_N}+\frac{\kappa_4}{4!}\frac{g_\sigma^2\sigma^2}{m_N^2}\Bigg)
\nonumber\\
&
+\frac{1}{2}m_{\omega}^2\,\omega^2
+\frac{1}{8}\zeta_{0}\,{g_{\omega}^2}\,\omega^4
+ \frac{1}{2}m_{\rho}^2\,\rho^{2}+3\Lambda_{\omega}\, g_\omega^2\, g_\rho^2\, \omega^2 \rho^2,
\label{eq:eden_rmf}
\end{align}
where $k_N$ and $m^*$ are Fermi momentum and effective mass ($m_N - g_\sigma\sigma$) of nucleons respectively. 

In addition to nucleons, NSs also contain a small amount of leptons, which are needed for charge-neutral stars. Therefore, the lepton Lagrangian density is given as \cite{Das_2021_fmode}
\begin{equation}
    {\cal L}_l = \sum_{l=e, \mu}\bar\psi_{l}\Big(i\gamma^{\mu}\partial_\mu-m_l\Big)\psi_l \, ,
\end{equation}
where $m_l$ is the mass of lepton ($m_e = 0.511$ MeV, and $m_\mu = 105.66$ MeV). The energy density of the leptons is given as
\begin{equation}
    \epsilon_l = \sum_l\frac{1}{\pi^2}\int_0^{k_l} dk \ \sqrt{k^2+m_l^2} \, .
    \label{eq:eden_l}
\end{equation}

\subsection{Neutron Decay DM Model}
\label{subsec:NDM}
Among three decay channels, the channel $n\rightarrow \chi + \phi$ (where $\chi$ and $\phi$ are dark fermions and light-dark bosons) is the most important one due to 2 solar mass constraints given by different NS observations. Like neutrinos, the light-dark boson will escape the NSs, and that sets the condition for chemical equilibrium ($\mu_n = \mu_\chi$) between neutron and DM. In this study, we chose the mass of the DM ($m_\chi$) equal to 938 MeV by following the Refs. \cite{Husain_2022a, Husain_2022b}.

The produced DM is treated as a degenerate Fermi gas with the self-interaction carried out by the exchange of vector bosons ($V$) having mass ($m_v$), and its coupling strength ($g_v$) as \cite{Shirke_2023}:
\begin{equation}
    \mathcal{L}_{\rm DM} =  \bar\chi(i\gamma_\mu \partial^\mu - m_\chi - g_v \gamma_\mu V^\mu)\chi -\frac{1}{4}V_{\mu\nu}V^{\mu\nu} + \frac{1}{2} m_v^2 V_\mu V^\mu \, .
\end{equation}
The energy density for the DM is given by 
\begin{equation}
    \epsilon_{\chi} = \frac{1}{\pi^2}\int_0^{k_{F_\chi}} k^2 \sqrt{k^2+m_\chi^2} \, dk + \frac{1}{2} \left(\frac{g_v}{m_v}\right)^2 n_\chi^2 \, ,
\end{equation}
where $k_{F_\chi}$ and $n_\chi$ are the DM Fermi momentum and number density, respectively. The chemical potential of the DM is given by 
\begin{equation}
    \mu_\chi = \sqrt{k_{F_\chi}^2 + m_\chi^2} + G_v n_\chi \, ,
\end{equation}
where $G_v$ is defined as $(g_v/m_v)^2$ and $n_\chi = \frac{k_{F_\chi}^3}{3\pi^2}$. Here, we neglect the DM-baryon interactions that could also stabilize NSs by making it energetically costly to produce DM particles in a pure baryonic medium, suppressing DM accumulation, and maintaining nearly pure NSs. \cite{Grienstein_2019}. Therefore, we plan to include this model in our future work.

The total energy density of the system is obtained by adding the contributions of nucleons, leptons, and DM, and it is given by $\epsilon = \epsilon_N + \epsilon_l + \epsilon_\chi$. The pressure can be calculated using the thermodynamical relation $p = \sum_i\mu_i n_i - \epsilon$, where $\mu_i$ and $n_i$ are the chemical potential and number density for nucleons, leptons, and DM, respectively, and their values are obtained by solving the following equations \cite{Husain_2022a, Husain_2022b, Shirke_2023, Routaray_2024}
\begin{align}
    \mu_n & = \mu_p + \mu_e \, , \nonumber \\
    \mu_\mu & = \mu_e \, , \nonumber \\ 
    \mu_\chi & = \mu_n \, , \nonumber \\
     n_p & = n_e + n_\mu \, , \nonumber \\
     n_b & = n_n + n_p + n_\chi \, .
     \label{eq:mu_den}
\end{align}
\subsection{Macroscopic Properties of NS}
\label{subsec:MPNS}
The macroscopic properties, such as mass and radius ($M-R$) of the NS, can be calculated by solving the Tolman-Oppenheimer-Volkoff equations \cite{TOV1, TOV2} for a given EOS. For the crust part, we adopt the Negele-Vautherin \cite{Negele_1973} EOS for the inner crust within the density range (0.001 fm$^{-3}$ < $\rho$ < $\rho_t$), and the Baym-Pethick-Sutherland \cite{Baym_1971} EOS for the outer crust ($\rho$ < 0.001 fm$^{-3}$). By ensuring a smooth transition in pressure and energy density between the crustal and core EOS branches, a transition density of approximately $\rho_t \approx 0.08$ fm${-3}$ is obtained. It is worth noting that the choice of crustal EOS does not affect the maximum mass of the NS, though it can have a slight impact on the radius and tidal deformability of NSs. The derived EOSs and their corresponding $M-R$ relations are shown in Fig. \ref{fig:eos_mr_tidal}, and this will be discussed in Sub-Sec. \ref{subsec:EMRT}. Once we have $M-R$ profiles, the other macroscopic properties, such as tidal deformability, moment of inertia, and oscillation properties, can be calculated by solving their corresponding equations as given in Refs. \cite{Hinderer_2010, Hartle_1968, Throne_1969}. 
\section{Result and Discussion}
\label{sec:rd}
\subsection{Coupling Constants in RMF Models}
\label{subsec:NMPs}
In this sub-section, we discuss the methodology used to generate the new models, which are discussed in detail in Refs. \cite{Chen_2014, Zhu_2023}. For a fixed RMF model, the unknown parameters in Eqs. (\ref{eq:rmf_lagrangian}, \ref{eq:eden_rmf}) are $g_\sigma$, $g_\omega$, $g_\rho$, $\kappa_3$, $\kappa_4$, $\zeta_0$, and $\Lambda_\omega$. These parameters can be obtained by fitting the nuclear matter (NM) properties, such as binding energy ($\epsilon_{\rm sat}$), incompressibility ($k_{\rm sat}$), symmetry energy ($j_{\rm sat}$), slope parameter ($l_{\rm sat}$), and nucleon effective mass ($m_{\rm sat}$) at saturation density ($n_{\rm sat}$). We calculate the values of all coupling constants by following the approach developed by \citet{Chen_2014} at saturation density except $\zeta_0$ because it has a significant impact on very high-density nuclear matter systems like NS. The value of $\zeta_0$ can be tuned to predict the soft/stiff behavior of the EOSs, which directly impacts the NS maximum mass. Therefore, in this study, we developed three models named ``HCD0, HCD1, and HCD2" for different values of $\zeta_0$, which satisfy the constraint of the NS maximum mass $\ge 2M_\odot$. The numerical values of coupling constants are provided in Table \ref{tab:NMP}, and they are fitted on the following chosen values of the NM saturation point, i.e. the saturation density $n_{\rm sat} = 0.15$ fm$^{-3}$, the binding energy $e_{\rm sat} = -16.0$ MeV, the nuclear compressibility $k_{\rm sat} = 250$ MeV, the symmetry energy $j_{\rm sat} = 33$ MeV and its first derivative $l_{\rm sat} = 66$ MeV, and the nucleon effective mass $m_{\rm sat} = 0.65$.

\begin{table}
\caption{RMF coupling constants, NS maximum mass ($M_{\rm max}$), canonical radius ($R_{1.4 M_\odot}$), and tidal deformability ($\Lambda_{1.4}$) are given.}
\label{tab:NMP}
\resizebox{\textwidth}{!}{%
\begin{tabular}{lllllllllllllll}
\hline
Model &
  $g_\sigma$ &
  $g_\omega$ &
  $g_\rho$ &
  $\kappa_3$ &
  $\kappa_4$ &
  $\zeta_0$ &
  $\Lambda_\omega$ &
  \begin{tabular}[c]{@{}l@{}}$M_{\rm max}$ \\ ($M_\odot$) \end{tabular} &
  \begin{tabular}[c]{@{}l@{}}$R_{1.4 M_\odot}$ \\ (km) \end{tabular} &
  $\Lambda_{1.4}$ &
  Astro &
  $\chi$EFT &
  Type \\ \hline
 HCD0 &  9.4544 &  11.7732 & 10.0442 & 1.8514 & -6.4302 & 0.0000 & 0.0225  & 2.52 & 13.42 & 748.93 & \checkmark & \checkmark  & stiff \\ 
 HCD1 &  9.5045 &  11.9267 & 10.1018 & 1.6127 & -3.7633 & 1.4225 & 0.0230  & 2.19 & 13.21 & 647.04 & \checkmark & \checkmark & inter. \\ 
 HCD2 &  9.5562 &  12.0865 & 10.1604 & 1.3716 & -1.0533 & 2.9216 & 0.0235  & 2.02 & 13.02 & 577.41 & \checkmark & \checkmark & soft \\ \hline
\end{tabular}%
}
\end{table}

In Table \ref{tab:NMP}, we also provide other crucial information on the three models developed here. The models are divided into three types: stiff, intermediate, and soft, according to the NS maximum mass by tuning the value of $\zeta_0$. All three models satisfy both the chiral effective field theory ($\chi$EFT) limit at the lower density ($0.5 < n_b/n_{\rm sat} < 1.5$) of pure neutron matter (PNM) \cite{Drischler_2016} and astrophysics (Astro) constraints, including NSs maximum mass ($\geq 2M_\odot$) from different pulsars, NICER measurements for mass and radius, i.e. $M = 1.34_{-0.15}^{+0.16} M_\odot, \, R = 12.71_{-1.19}^{+1.14}$ km for PSR J0030+0451 \cite{Riley_2019} and $M = 2.072_{-0.066}^{+0.067} M_\odot, \, R = 12.39_{-0.98}^{+1.30}$ km for PSR J0740+6620 \cite{Riley_2021}, and the loosely bound tidal deformability limit from GW170817 event ($\Lambda_{1.4}<800$) \cite{GW170817_2017}. The detailed discussions are explained in the following sub-sections.
\begin{figure}
    \includegraphics[width=0.9\textwidth]{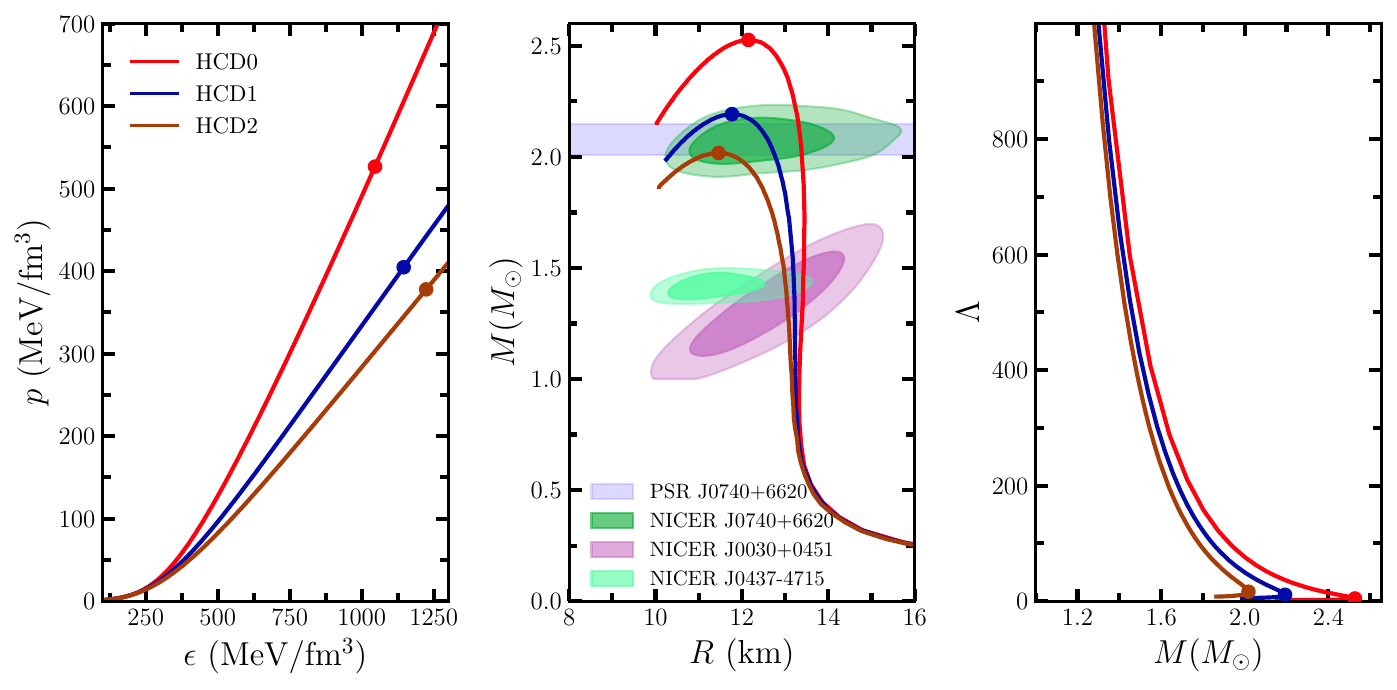}
    \caption{{EOSs}(left), $M-R$ relations (middle), and tidal deformability (right) are shown for HCD0-2 models.  The marker $\CIRCLE$ represents the value corresponding to the maximum mass cases. The contours of 68 and 95 \% are from three separate NICER observations for different pulsars \cite{Riley_2019, Riley_2021, Choudhury_2024}. The horizontal fill band represents the maximum mass constraints from Fonseca {\it et al.} \cite{Fonseca_2021} for PSR J0740+6620.}
    \label{fig:eos_mr_tidal}
\end{figure}
\subsection{EOSs, $M-R$ Relations, and Tidal Deformability}
\label{subsec:EMRT}
The EOS of the NS is calculated by solving Eqs. (\ref{eq:rmf_lagrangian}--\ref{eq:eden_l}) imposing $\beta$-equilibrium and charge neutrality condition given in Eq. (\ref{eq:mu_den}), without DM. The calculated EOSs, $M-R$ relations, and tidal deformability are shown in the left, middle, and right panels of Fig. \ref{fig:eos_mr_tidal} for newly developed models. From HCD0-2, the nature of the EOS is stiff to soft due to the absence/presence of the self-interaction of the $\omega$-coupling term. In the $M-R$ plot, maximum masses and their corresponding radii are shown with a circle marker. HCD0, HCD1, and HCD2 predict the maximum masses $2.52, 2.19$, and 2.02 $M_\odot$ which perfectly pass in the observational bands given by Fonseca {\it et al.} \cite{Fonseca_2021} $\left(M = 2.08\pm 0.07 M_\odot \right)$ respectively. The simultaneous mass-radius constraints provided by NICER are consistent for the three models, as shown in the figure. In the case of tidal deformability, the values are 748.93, 647.04, and 577.41 for HCD0-HCD2 for canonical stars, which satisfy the GW170817 limit. 
\subsection{Effects of NDA on Properties of NSs}
\label{subsec:NDA_NS_Props}
For these newly developed models, we want to calculate the effects of NDA on the EOSs and various properties of the DMANS. Therefore, we choose $G_v$ as a free parameter in this model, and we fix the DM mass to be equal to 938 MeV.

\begin{figure}
    \includegraphics[width=0.65\textwidth]{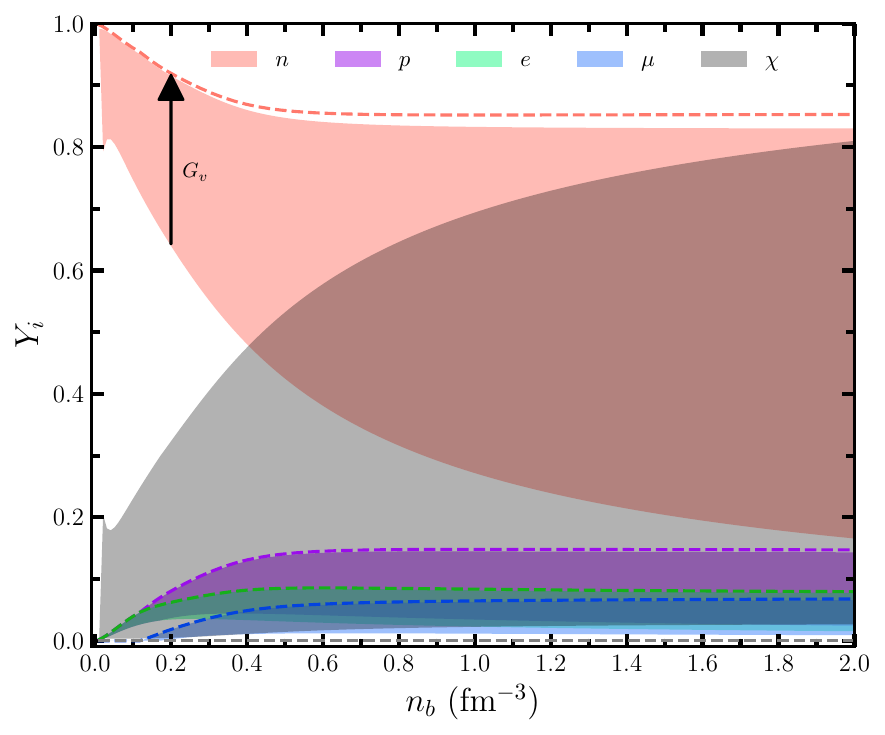}
    \caption{Particle fractions for all species, including DM, are shown as a function of baryon density for the HCD0 model. The shaded areas are calculated for different $G_v$ values in the range $0-300$ fm$^2$. The dashed colored lines represent the composition inside the NSs without DM. The vertical arrow represents the trend of the neutron population with increasing $G_v$ values.}
    \label{fig:pf_den_dm}
\end{figure}

First, we want to calculate the particle fraction (PF) of the different species to showcase the amount of DM formation inside the NSs due to neutron decay with different self-interaction strengths. For a fixed value of $G_v$, we solve Eq. (\ref{eq:mu_den}) for total baryon density up to 2 fm$^{-3}$. The obtained PFs for the different species are shown in Fig. \ref{fig:pf_den_dm} for the HCD0 model with the values of $G_v$ from $0-300$ fm$^2$ to observe the effects of DM inside the star. It is observed that DM content is maximal inside the star if there is no self-interaction ($G_v=0$) between them, and it decreases to almost zero with an increasing value of $G_v$. Other PFs are following the opposite trend compared to DM. Similar predictions have also been discussed in other studies \cite{Husain_2022a, Husain_2022b, Shirke_2023, Routaray_2024}.

\begin{figure}
    \includegraphics[width=0.9\textwidth]{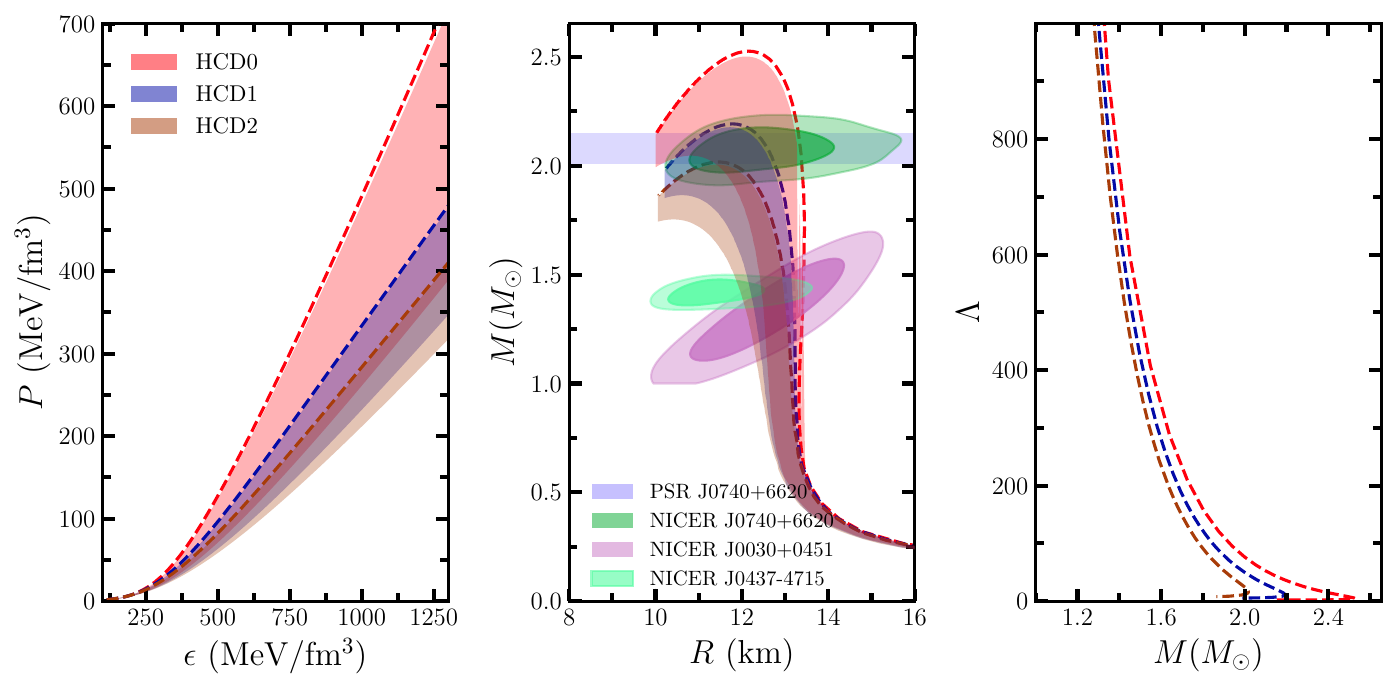}
    \caption{EOSs (left), $M-R$ relations (central), and tidal deformability (right) panels are shown for three models HCD0-2 by varying $G_v$ values, ranging from 10 to 300 fm$^2$. The horizontal filled band represents the maximum mass constraints from Fonseca {\it et al.} \cite{Fonseca_2021} for PSR J0740+6620. The contours of 68 and 95 \% are from three separate NICER observations for different pulsars as mentioned in the legend \cite{Riley_2019, Riley_2021, Choudhury_2024}.}
    \label{fig:eos_mr_tidal_dm}
\end{figure}

In Fig.\ref{fig:eos_mr_tidal_dm}, we show the EOSs, $M-R$ profiles, and tidal deformability in the three panels for three models HCD0-HCD2 with a similar range of $G_v$ values. Due to the production of DM inside the star, the EOS of the DMANS becomes softer, which directly affects the magnitude of the macroscopic properties. Without self-interactions ($G_v=0$), the DM content is too high, which provides the softest EOS and produces a maximum mass around $\sim 0.7 M_\odot$ as mentioned in Refs. \cite{Baym_2018, Mckeen_2018, Husain_2022a}. Configurations characterized by low values of $G_v$ are not shown in Fig. \ref{fig:eos_mr_tidal_dm}, and they are actually neglected due to the $2 M_\odot$ constraint imposed by several pulsar maximum masses. A detailed discussion of the range of $G_v$ is done in the following subsections based on other observational aspects.

The magnitude of mass and radius decreases with decreasing $G_v$ due to softened EOSs. For example, the maximum mass and its corresponding radius with $G_v=300$ fm$^2$ for the HCD0 model are $2.50 M_\odot$ and 12.1 km, and their values reduce to $2.05 M_\odot$ with a radius of 10.82 km for $G_v = 10$ fm$^2$. The softness is also strong in other models, such as HCD1 and HCD2. Therefore, to constrain the model dependency for different values of $G_v$, we put observational data such as NICER and different pulsars on top of the $M-R$ relations plot. If we stick to the maximum mass limit by different pulsars, only HCD0 and HCD1 models up to some values $G_v$ will satisfy the $2M_\odot$ limit. However, the NICER constraints are well reproduced by all three models with all values of $G_v$.

The tidal deformability ($\Lambda$) is another crucial quantity used to constrain the unknown parameters of this model because its value has already been measured for the binary NSs merger in the GW170817 event \cite{GW170817_2017} $\left(\Lambda_{1.4}< 800\right)$. Also, in this case, $\Lambda$ follows the same trend compared to $M-R$ relations. For example, in the HCD0 model, the values of $\Lambda_{1.4}$ decrease from 748.93 (without DM) to 435.81 (with DM and $G_v=10$ fm$^2$). It is observed that with the production of a finite amount of DM inside the NS due to neutron decay, the value of $\Lambda_{1.4}$ is consistent with the GW170817 limit. Therefore, tidal deformability can be taken as an additional tool to constrain the DM fraction (DMF) inside the NSs.

The exact content of DM can be calculated as 
\cite{Shirke_2023, Shirke_2024}
\begin{equation}
    f_\chi = \frac{M_\chi}{M} = \frac{\int_0^R \epsilon_\chi \, d^3r}{\int_0^R \epsilon \, d^3r} \, ,
\end{equation}
where $M_\chi$ ($\epsilon_\chi$) and $M$ ($\epsilon$) are the mass (energy density) of the DM component and the total one, respectively. For HCD0-2 models, we find $f_\chi^{\rm max} = 0.24, 0.20$, and $0.18$ respectively for maximum mass configuration with $G_v=10$ fm$^2$. For higher values of $G_v$, the DM fraction decreases due to the increase of the self-interaction between DM, as discussed above. Therefore, we can constrain the DM fraction as well as the self-interaction strength with the help of several observational data, such as the maximum mass limit from several pulsars, the simultaneous $M-R$ data from NICER, and the tidal deformability limit from GW observation. Another way to constrain the parameter $G_v$ and DM fraction is by utilizing NS cooling observational data, which will be discussed in a future study.
\begin{figure*}
    \includegraphics[width=0.8\textwidth]{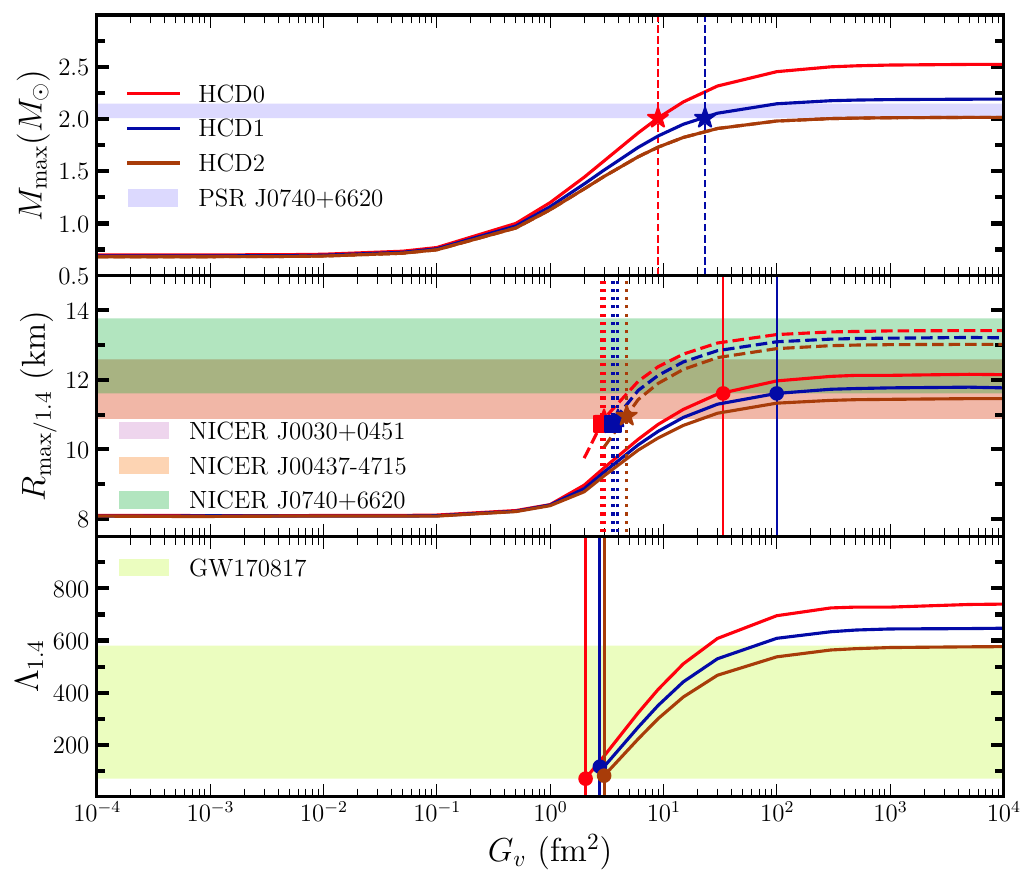}
    \caption{Maximum mass ($M_{\rm max}$), radius corresponds to $M_{\rm max/1.4}$, and canonical tidal deformability are shown as a function of $G_v$. Solid and dashed lines are for different observations. Similarly, the markers $\CIRCLE$, $\blacksquare$ and $\bigstar$ represent the lower bound of $G_v$ obeying different observational data.}
    \label{fig:mrl_gv_dm}
\end{figure*}
\subsection{Constraints on $G_v$ from NS Observations}
\label{subsec:CON_Gv}
In this study, the unknown parameter of the DM is $G_v$, which directly affects the DM fraction inside the star. The $G_v$ value can be constrained with the help of NS observables  recently measured by different observations, i.e. 
\begin{enumerate}
    \item Maximum mass limit imposed by the PSR J0740+6620 having mass $M = 2.08 \pm 0.07 M_\odot$ \cite{Fonseca_2021}.
    \item Radius constraints by NICER for PSR J0030+0451 \cite{Vinciguerra_2024}, J0437-4715 \cite{Choudhury_2024}, and J0740+6620 \cite{Salmi_2022} are $ R_{1.4 M_\odot} = 11.71_{-0.83}^{+0.88}$ km, $ R_{1.418 M_\odot} = 11.36_{-0.63}^{+0.95}$ km, and $R_{2.07 M_\odot}=12.49_{-0.88}^{+1.28}$ km respectively.
    \item Tidal deformability constraints by GW170817 event \cite{GW170817_2017}, $\Lambda_{1.4 M_\odot} = 190_{-120}^{+390}$\,.
\end{enumerate}

In Fig.\ref{fig:mrl_gv_dm}, we show the maximum mass $M_{\rm max}$ (upper panel), the radius corresponding to the maximum and canonical mass configurations (middle panel), and the tidal deformability for canonical stars (lower panel) with $G_v$ values in the range $10^{-4}-10^4$ fm$^2$. The above-mentioned observational constraints are overlaid to check the model consistency by varying $G_v$. The lower bounds of $G_v$ are also shown with different markers. In case of $M_{\rm max} \sim G_v$, HCD0 (HCD1) have lower bounds around 8.95 (23.30)  fm$^2$ for PSR J0740+6620. In the case of HCD2, its maximum mass is around 2.02, which just touches the lower value of PSR J0740+6620, meaning that soft EOSs are not useful in this scenario. 

In the case of $R_{\rm max, 1.4} \sim G_v$, all three models (HCD0-2) provide the lower values of $G_v$ and are 3.0, 3.96, and 4.75 fm$^2$, respectively for NICER(J0030+0451) and these values are almost similar in case of NICER(J0437-4715). However, for NICER(J0740+6620), only HCD0 and HCD1 provide the lower cut-off of 33.68 and 100 fm$^2$, respectively. If we stick to the tidal deformability constraints, again, all three models provide similar lower limits ($2-5$ fm$^2$), as shown in the figure. Hence, we observe that different observational results provide different lower values for the self-interacting strengths. Therefore, we also need galaxy cluster data to put joint constraints on $G_v$. To check the consistency of our findings, we have also included several established RMF EoSs in our analysis, such as BigApple \cite{Fattoyev_2020, Das_BA_2021}, HTZCS \cite{Hornick_2018}, NITR1 \cite{Routaray_NITR1_2024}, IOPB-I \cite{Kumar_2018, Parmar_2022}, and FSUGarnet \cite{Chen_2015}. The results obtained using these models exhibit similar trends to those seen with our HCD models.
\subsection{Constraints on $G_v$ and DM Fraction with Galaxy Cluster Data}
In addition to NS observables, the scattering cross-section limits from different galaxy clusters can impose constraints on the self-interacting strength. The scattering cross-section within the Born approximation at a small velocity limit is given as \cite{Girmohanta_2022}:
\begin{align}
    \sigma & \approx \frac{4\pi\alpha_\chi^2m_\chi^2}{2 m_v^4}  = \frac{G_v^2 m_\chi^2}{8\pi}
    \, , \nonumber \\
    \Rightarrow \sigma & \approx 2\pi \left(\frac{\alpha_\chi}{0.01}\right)^2 \left(\frac{m_\chi}{10\, {\rm GeV}}\right)^2 \left(\frac{10 \, {\rm MeV}}{m_v}\right)^4  \frac{(0.01)^2 (10 \, {\rm GeV})^2}{(10 \, {\rm MeV})^4}
    \, , \nonumber \\
    \Rightarrow \sigma &\approx \frac{2\pi \times 10 ^{-4} \times 10^{2} \,  {\rm GeV^2}}{10^4 \, {\rm MeV}^4} \left(\frac{\alpha_\chi}{0.01}\right)^2 \left(\frac{m_\chi}{10 \, {\rm GeV}}\right)^2 \left(\frac{10 \, {\rm MeV}}{m_v}\right)^4
    \,, \nonumber \\
    \Rightarrow \sigma &\approx \frac{2\pi \times 10^{-6} \, {\rm GeV^2}}{{\rm MeV}^4} \left(\frac{\alpha_\chi}{0.01}\right)^2 \left(\frac{m_\chi}{10 \, {\rm GeV}}\right)^2 \left(\frac{10 \, {\rm MeV}}{m_v}\right)^4
    \,, \nonumber \\
    \Rightarrow \sigma &\approx 2.5 \times 10^{-21} \left(\frac{\alpha_\chi}{0.01}\right)^2 \left(\frac{m_\chi}{10 \, {\rm GeV}}\right)^2 \left(\frac{10 \, {\rm MeV}}{m_v}\right)^4 \,  {\rm cm}^2 
    \,, \nonumber \\
    \Rightarrow \sigma & \approx 2.5 \times 10^{-21} \left(\frac{\alpha_\chi}{0.01}\right)^2 \left(\frac{m_\chi}{10 \, {\rm GeV}}\right)^2 \left(\frac{10 \, {\rm MeV}}{m_v}\right)^4 \,  {\rm cm}^2 ,
    \label{eq:sigma_chi}
\end{align}
where we use $1 \, {\rm GeV} = 5.06 \times 10^{13} \, {\rm cm}^{-1}$, $1 \,  {\rm MeV} = 5.06 \times 10^{10} \, {\rm cm}^{-1}$. This expression is two orders of magnitude larger than the expression given in Refs. \cite{Tulin_PRL_2013, Tulin_PRD_2013} \endnote{Typo in equation-1 of that Phys. Rev. Lett. paper is confirmed by \citet{Tulin_PRL_2013}, and the correct one is given in equation 3 of this paper \cite{Tulin_PR_2018}}. The $\alpha_\chi$ is the fine structure constant, having a value of $g_v^2/4\pi$, which can be expressed in terms of $G_v (= g_v^2/m_v^2)$ as $\alpha_\chi = G_v m_v^2/4\pi$. Now, we want to express Eq. (\ref{eq:sigma_chi}) in terms of $G_v$ as follows \cite{Shirke_2023}
\begin{figure}
    \includegraphics[width=0.8\textwidth]{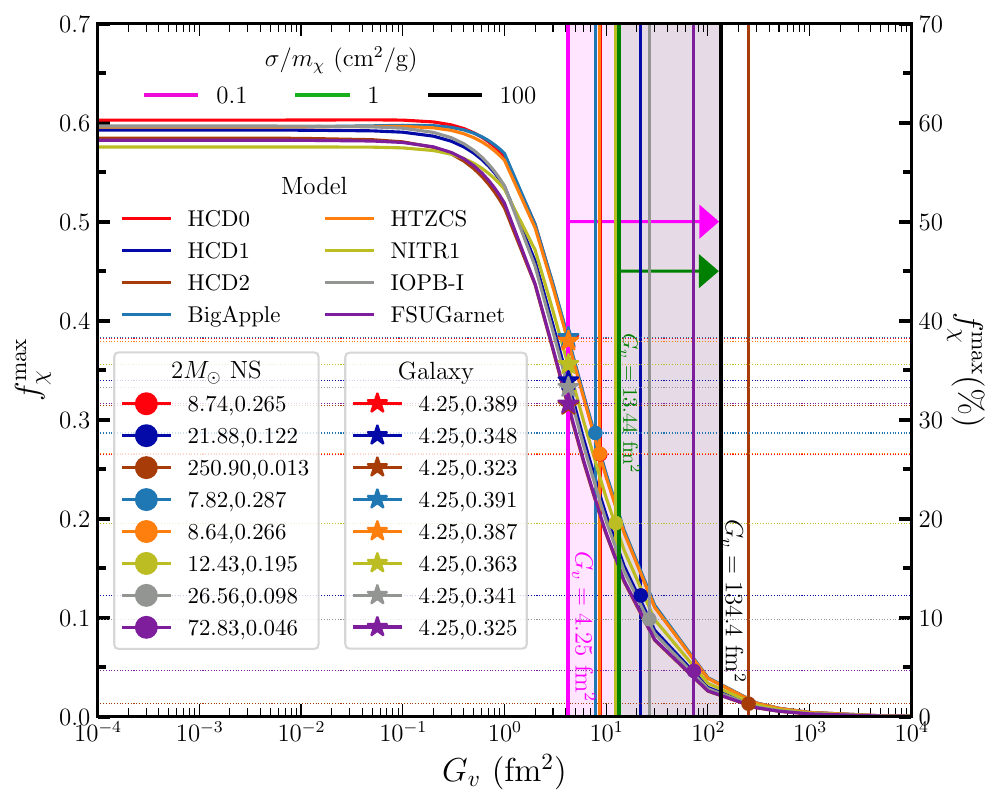}
    \caption{DM fractions with different values of $G_v$ using Eq. \eqref{eq:sigma_chi}. Markers $\CIRCLE$ and $\bigstar$ represent the lower bound of $G_v$ satisfying the $2 M_\odot$ constraint and Galaxy cluster limit, respectively. The fill areas (red, blue, and rust) represent the values of $G_v$ for which different models achieve $2 M_\odot$ limit, and the magenta-fill area represents the bounds provided by the galaxy cluster. The vertical magenta, green, and black lines are for $G_v$ values corresponding to different $\sigma/m_\chi$ values.}
    \label{fig:frac_gv_dm}
\end{figure}
\begin{align}
   \frac{\sigma}{m_\chi} & \approx 0.59 \times 10^{-2} \left(\frac{G_v}{1 \, {\rm fm}^2}\right)^2 \left(\frac{m_\chi}{1 \,  {\rm GeV}}\right) \, \frac{{\rm cm}^2}{{\rm g}} \, ,\\
   G_v & = 13.44 \, \sqrt{\left(\frac{\sigma}{m_\chi}\right) \left(\frac{{\rm g}}{\rm cm^2}\right)} \, \, {\rm fm^2}, \, {\rm for} \, m_\chi=938 \, {\rm MeV} .
   \label{eq:sigma_mchi_gv}
\end{align}
\begin{itemize}
    \item From galaxy clusters \cite{Loeb_2011, Manoj_2016, Sagunski_2021}, $\frac{\sigma}{m_\chi} = 0.1 \, \frac{\rm cm^2}{{\rm g}}$, \, $G_v = 4.25$ fm$^2$,

    \item For Bullet cluster \cite{Randall_2008}, $\frac{\sigma}{m_\chi} < 1 \, \frac{\rm cm^2}{{\rm g}}$, \, $G_v < 13.44$ fm$^2$,
    
    \item For dwarf galaxy \cite{Oh_2011}, $\frac{\sigma}{m_\chi} = 1 \, \frac{\rm cm^2}{{\rm g}}$, \, $G_v = 13.44$ fm$^2$,

    \item From core-cusp problem \cite{Manoj_2016, Sagunski_2021}, $\frac{\sigma}{m_\chi} \lesssim 100 \, \frac{\rm cm^2}{{\rm g}}$, \, $G_v \lesssim 134.4$ fm$^2$,
\end{itemize}

In Fig.\ref{fig:frac_gv_dm}, we show the DMF as a function of $G_v$ for the three models. The lower bound of $G_v$ is 4.25 fm$^2$ as mentioned above, which provides $f_\chi \lesssim 38.9, 34.8,$ and 32.3\% for HCD0-2 models. If we stick only to the $2 M_\odot$ constraint, the lower bounds of $G_v$ (DMF) are 8.74 (26.5\%), 21.88 (12.2\%), and 250.90 (1.3\%) fm$^2$ for HCD0-2 models, respectively, as reported also in the upper panel of Fig.\ref{fig:mrl_gv_dm} for maximum mass scenario. The higher limit of $G_v$ is $\sim 134.4$ fm$^2$, which reduces the DMF. 

Hence, the minimum value of $G_v$ from galaxy observational data allows the highest DMF $\sim 39\%$, and NS observation allows $\sim 27\%$ for the HCD0 type model. Other EOSs are also predicted to have the highest DMF between 32-39\% for the lower bound of $G_v$. The softest EOS, like HCD2, is completely out of the boundary region provided by Galaxy clusters. Therefore, to constrain the unknown nature of $G_v$ and DMF, we need a good statistical analysis tool, such as Bayesian inference or machine learning, to exactly estimate the lower bounds of $G_v$, which we will explore in the future.

From the above discussion, we observe that the lower bound of $G_v=4.25$ fm$^2$ provides information on the minimum strength of self-interaction between DM based on the galaxy clusters data $\left(\frac{\sigma}{m_\chi} = 0.1 \, \frac{\rm cm^2}{{\rm g}}\right)$. However, this value is inconsistent with NS mass observation, which requires the maximum mass to be $>2 M_\odot$, yielding different values of $G_v$ depending on the model used. For example, $G_v=4.25$ fm$^2$ corresponds to a DMF $\sim 39\%$ inside the NS, which doesn't align with the $2 M_\odot$ limit. Nevertheless, when comparing the $G_v$ values associated with either the Bullet Cluster or dwarf galaxy data $\left(\frac{\sigma}{m_\chi} \leq 1 \, \frac{\rm cm^2}{{\rm g}}\right)$ to the stiffest EOSs like HCD0, BigApple and HTZCS, their $G_v$ values are almost similar and a little difference in DMF, thus predicting DMF results compatible with the $2 M_\odot$ limit. A similar trend is also followed by moderately stiff EOSs such as HCD1 and IOPB-I. The upper bound $G_v=134.4$ fm$^2$ is also important, as it helps to rule out the softest EOS-like HCD2 model considered in this study. However, other EOSs predict similar values of $G_v$, and it depends on the stiffness nature of the EOS. Therefore, the scattering cross-section data play a crucial role in constraining the strength of self-interaction between DM particles.
\section{Conclusions}
In this study, we investigate the effects of DM on various macroscopic properties of NSs based on the neutron decay anomaly model. For the nucleonic case, we construct three new models, HCD0–HCD2, within the RMF framework, with NM saturation values consistent with experimental and empirical constraints established in our previous study \cite{Burgio_2024}. The coupling constants for the three models are determined by fitting the NM properties at saturation density. EOS for PNM also conforms to the $\chi$EFT band in the lower density regime. For NSs, the EOSs are derived under the conditions of $\beta$-equilibrium and charge neutrality. The macroscopic properties predicted by these three models satisfy key observational constraints. Specifically, the maximum NS mass exceeds the $2M_\odot$ limit, while the radius predictions align with NICER's $M$-$R$ constraints. Additionally, the calculated tidal deformability values are consistent with the constraints from the GW170817 event.

The DM EOS is calculated using the NDA model, where neutrons decay to a few DM. The self-interaction between DM is also considered in this case by the exchange of the vector bosons. The DM PF increases with decreasing the strength of the self-interaction parameter. The EOSs for DMANS are getting softer with a lower value of $G_v$ compared to a higher one. It signifies that the lower the value of self-interaction strength, the more DM content is inside the star, which softens the EOS and decreases the magnitude of mass, radius, and tidal deformability magnitude. With the help of NS observational data, we can impose constraints on the DM unknown parameters. 

We calculate the mass, radius, and tidal deformability for the three models using different $G_v$ values, combining observational data to determine the range of $G_v$ that satisfies these constraints. Our analysis shows that only the HCD0 model meets both the maximum mass limit and NICER data, while the other two models align with NICER but do not fully comply with the maximum mass constraints for all values of $G_v$ considered in this study. Additionally, we compute $M_{\rm max}$, $R_{\rm max/1.4}$, and $\Lambda_{1.4}$ for the HCD0-2 models by varying $G_v$ from $10^{-4}$ to $10^4$ fm$^2$. The results indicate that $G_v$ must exceed a certain lower bound to satisfy constraints imposed by pulsar mass measurements, NICER, and GW observations. Notably, except for HCD2, the other two models exhibit a lower bound on $G_v$ for the maximum mass requirement. Regarding the radius for canonical neutron stars, all three models satisfy NICER constraints for J0030+0451, while only HCD0 and HCD1 meet the limits for J0740+6620. In the case of $\Lambda_{1.4}$, all models predict lower cutoff values consistent with GW data.

Finally, we impose qualitative constraints on $G_v$ and the DM fraction using data from different galaxy clusters. Our analysis reveals that the minimum value of $G_v$ is 4.25 fm$^2$, based on the scattering cross-section over the DM mass limit imposed by galaxy clusters. The corresponding DM fraction are found to be $\sim 39\%$ for the stiffest EOSs such as HCD0, BigApple and HTZCS considered in this study. However, if we consider only the constraint that the maximum mass must exceed $2M_\odot$, the lower bounds on $G_v$'s are relaxed to 8.74, 8.64, and 7.82  fm$^2$, corresponding to a higher DM fraction of 26.5, 28.7, and 26.6\% for HCD0, BigApple, and HTZCS models, respectively. Therefore, in order to find the lower and upper bounds for the self-interacting strength, the Galaxy cluster data are crucial. However, for the precise determination of the values of $G_v$ and $f_{\chi}$, a robust statistical approach such as Bayesian inference or machine learning is required. This would help to constrain their parameter space using a combination of NS observational data and galaxy cluster cross-section measurements. We hope that future studies will explore this direction to obtain a more precise estimation of these parameters.
\vspace{6pt} 

\authorcontributions{Conceptualization, H. C. D. and G. F. B.; data curation, H. C. D.; formal analysis, H.C.D.; investigation, H.C.D., and G. F. B.; writing—original draft, H. C. D.; writing—review and editing, G. F. B. All authors have read and agreed to the published version of the manuscript.}

\funding{This research received no external funding.}

\dataavailability{The data presented in this study are available on request from the corresponding author.} 

\acknowledgments{H.C.D. would like to thank Hans-Josef Schulze for the discussions on the scattering cross-sections of the dark matter.}

\conflictsofinterest{The authors declare no conflicts of interest.}

\begin{adjustwidth}{-\extralength}{0cm}
\printendnotes[custom] 

\reftitle{References}

\bibliography{nda.bib}

\PublishersNote{}
\end{adjustwidth}
\end{document}